\documentclass[12pt]{article}
\usepackage[english]{babel}
\usepackage{graphicx}
\usepackage{amssymb}
\usepackage{natbib}
\usepackage{titling}
\usepackage[margin=2cm]{geometry}

\begin{document}

\title{The atmospheric fragmentation of the 1908 Tunguska Cosmic Body: reconsidering the possibility of a ground impact}
\author{L. Foschini\thanks{INAF Osservatorio Astronomico di Brera, Via Bianchi 46, 23807, Merate (Italy). Email: \texttt{luigi.foschini@inaf.it}.}, L. Gasperini\thanks{CNR Istituto di Scienze Marine, Via Gobetti 101, 40129, Bologna (Italy)}, C. Stanghellini\thanks{INAF Istituto di Radioastronomia, Via Gobetti 101, 40129, Bologna (Italy)}, R. Serra\thanks{Dipartimento di Fisica e Astronomia, Universit\`a di Bologna, Via Irnerio 46, 40126, Bologna (Italy)}, A. Polonia$^2$, G. Stanghellini$^2$}
\date{5 June 2018}
\thanksmarkseries{arabic}
\maketitle

\begin{abstract}
The 1908 June 30 Tunguska Event (TE) is one of the best studied cases of cosmic body impacting the Earth with global effects. However, still today, significant doubts are casted on the different proposed event reconstructions, because of shortage of reliable information and uncertainties of available data. In the present work, we would like to revisit the atmospheric fragmentation of the Tunguska Cosmic Body (TCB) by taking into account the possibility that a metre-sized fragment could cause the formation of the Lake Cheko, located at about $9$~km North-West from the epicentre. We performed order-of-magnitude calculations by using the classical single-body theory for the atmospheric dynamics of comets/asteroids, with the addition of the fragmentation conditions by \cite{FOSCHINI3}. We calibrated the numerical model by using the data of the Chelyabinsk Event (CE) of 2013 February 15. Our work favours the hypothesis that the TCB could have been a rubble-pile asteroid composed by boulders with very different materials with different mechanical strengths, density, and porosity. Before the impact, a close encounter with the Earth stripped at least one boulder, which fell aside the main body and excavated the Lake Cheko. We exclude the hypothesis of a single compact asteroid ejecting a metre-sized fragment during, or shortly before, the airburst, because there is no suitable combination of boulder mass and lateral velocity.
\end{abstract}

\section{Introduction}

The 1908 Tunguska Event (TE) has attracted the scientific curiosity of many researchers for more than one century. After the earliest studies, it was evident that the $2150$~km$^{2}$ wide devastation of the Siberian taig\`a was the result of the impact of a cosmic body with the Earth (see \citealt{VASILYEV}, \citealt{LONGO1}, and \citealt{ARTESHU} for reviews). However, there was a strong disagreement about its nature: comet or asteroid? 

In the latest twenty years of debate, the main novelty was the hypothesis proposed by \citet{GASP1} that the Lake Cheko could be an impact crater generated by a major fragment of the Tunguska Cosmic Body (TCB). This was one the results of the \emph{Tunguska99} scientific expedition, led by Giuseppe Longo (University of Bologna, Italy; \citealt{TUNGUSKA99}). In 1999 July, the \emph{Tunguska99} scientists and engineers explored the Lake Cheko and its surroundings, which are placed at about $9$~km North-West from the epicentre (azimuth\footnote{Clockwise from North.} $\sim 348^{\circ}$ from the Fast coordinates, \citealt{LONGO2}; see Fig.\ref{fig:map}). The reason for selecting Lake Cheko was that it could be an efficient collector for microparticles of the TCB. However, as measurements and observations went ahead, the hypothesis of an impact crater  emerged. More expeditions preceded and followed \emph{Tunguska99} in 1991, 1998, 2002, 2008, and 2009, to collect more data\footnote{See \texttt{http://www-th.bo.infn.it/tunguska/} for details.}. The main findings of these studies are: 

\begin{figure}[!t]
\begin{center}
\includegraphics[scale=0.8]{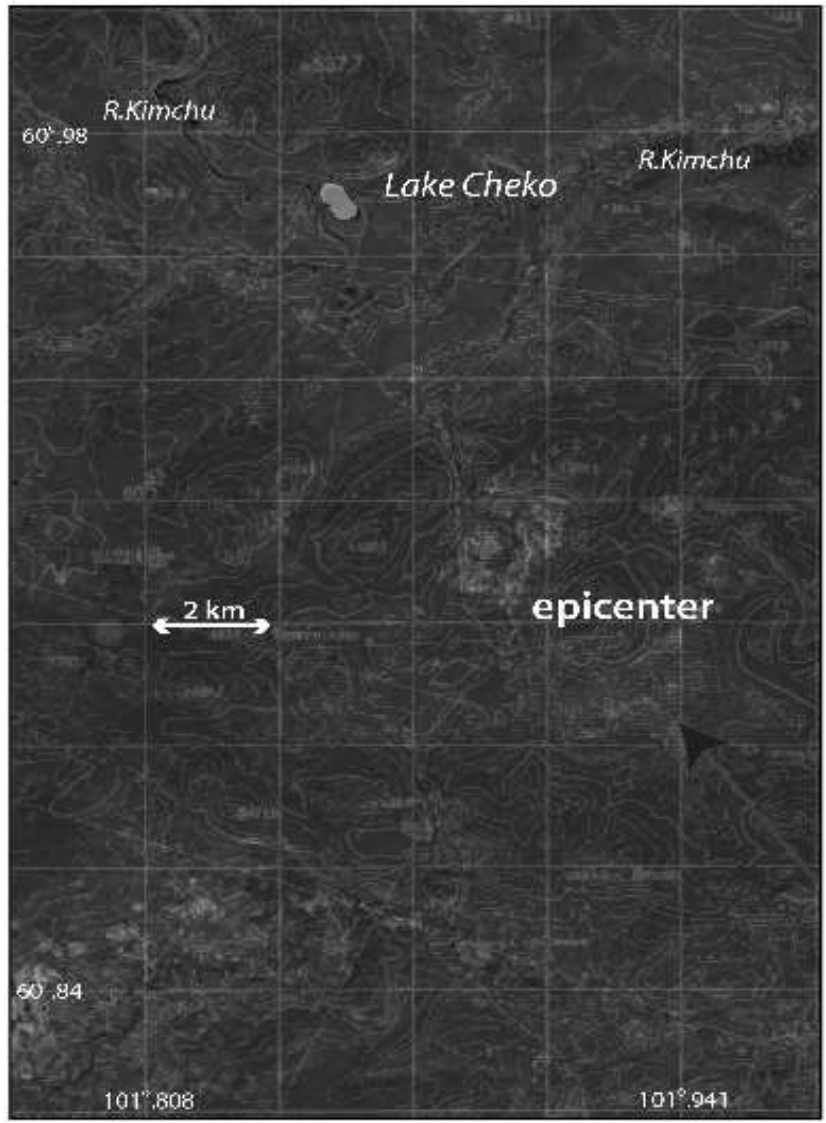}
\caption{Topographic map of the TE region over a Google Earth image. Location of Lake Cheko and approximate position of the airburst epicentre \citep{LONGO2} are indicated, as well as a likely azimuth of the TCB trajectory (adapted from \citealt{GASP6}).}
\label{fig:map}
\end{center}
\end{figure}

\begin{itemize}
\item {\it Lake Shape:} Lake Cheko has a cone-like shape, with ellipsoidal base ($\sim490\times 340$~m) and $\sim 50$~m depth; it is different from other local thermokarst lake, which in turn are much more shallow. By excluding the shallower parts ($\lesssim2$~m depth), the lake surface is almost circular with a diameter of $\sim 350$~m  \citep{GASP1,GASP6}.

\item {\it Sediment Cores:} Sediment cores collected in the lake show an upper part ($\sim 80$~cm long in the case of TG22) with laminated sediments, a transition section $\sim 40$~cm long, and a lower chaotic/massive part ($\sim 55$~cm long); $^{210}\mathrm{Pb}$ and $^{137}\mathrm{Cs}$ dating set the transition epoch around $1908$ \citep{GASP1,GASP2,GASP3}.

\item {\it Pollens:} Pollens found in the cores indicate the presence of aquatic plants, which are currently colonising the lake, only in the upper part, above the transition section \citep{GASP3}.

\item {\it Lake Bottom:} The laminated sediment covering the bottom of the lake is merged with fallen trees. Branches are clearly emerging from the mud. This supports the hypothesis that the lake is covering an old forest \citep{GASP5}.

\item {\it Dendrochronology:} Dendrochronological analysis of the trees around the lake indicates an increased ring growth after an epoch consistent with 1908, similarly to what is found in the other part of the devastation area. This was likely caused by the sudden availability of light for the survived plants and lack of competition (many trees, low light, low growth; less trees, high light, high growth). If the lake was already present before $1908$, we should expect no changes in the rings size of the lake trees, as light and nutrients would have been provided by the presence of the lake itself and the water. The small rings before 1908 in the survived trees around the lake are consistent with the presence of a forest in place of the lake before the TE. It is worth noting that during the latest $20-30$ years, the rings of the trees in the devastation area were back to the pre$-1908$ size, as there is again competition. Instead, the trees on the lake shores do not show any size change \citep{GASP5,FANTUCCI}.

\item {\it Fallen Trees Pattern:} The patterns of tree fall in the area destroyed by the TE are compatible with one or two airbursts. The disintegration in the atmosphere of its main part caused the forest devastations, and could have had a braking effect on the minor body protecting it from fragmentation \citep{LONGO2,LONGO3}.

\item {\it Trees Rings:} Partial or full tree defoliation caused by the TE has led to the formation of tree rings less dense, less coloured and more friable than older rings. This phenomenon is present not only in larches, but also in each species of conifers (\emph{Picea obovata}, \emph{Pinus cembra}, and \emph{Larix sibirica}) in the Tunguska area  \citep{LONGO4}.

\item {\it Peat Samples:} Analysis of peat samples collected close to the lake show an increase of $^{15}\mathrm{N}$ in an epoch consistent with $1908$, likely due to acid rains caused by an increase of particles suspended into the atmosphere \citep{KOLESNIKOV,TOSITTI}.

\item {\it Lake Bottom Anomalies:} Magnetometric and seismic reflection data collected from the lake surface indicate the presence of density and magnetic anomalies at the center of the lake, below $\sim 10$~m of sediments. These anomalies could be interpreted as a buried metre-sized body \citep{GASP4}.

\item {\it Geographical maps:} There is neither historical nor geographical indication that the lake existed before 1908: available maps of the region, including a 1883 topographic map prepared by the Russian Army, does not show any sign of the lake \citep{GASP1,LONGO3}. 
\end{itemize}

All these works, particularly the analysis of core sediments and survived trees close to the lake shores, point to the formation of the Lake Cheko in an epoch consistent with the TE. It is worth noting that there is no clear-cut evidence supporting the impact crater hypothesis, no smoking gun, but there are many observations and experimental results all consistent with the hypothesis that the Lake Cheko was generated at the epoch of the TE. 

This reconstruction was challenged by \citet{COLLINS}, and \citet{ROGOZIN}. The latter concluded that Lake Cheko is older than 1908 on the basis of $^{137}\mathrm{Cs}$ and $^{210}\mathrm{Pb}$ dating of two sediment cores. However, these authors rested their conclusions on radionuclides age estimate only, which is reliable for a limited time span after the 1908, and extrapolated their observation back in time by using a linear model. This assumption is not grounded as in 2017 we were already beyond the reliability of dating with the above cited isotopes (cf \citealt{GASP3}). 

The main objection raised by \citet{COLLINS} was about the lake morphology, whose depth-to-diameter ratio ($50/350 \sim 0.14$) is too small for typical known craters, and the lack of a rim. \citet{GASP2} replied that those characteristics could be explained if we consider a soft impact on wet target, where the transient crater is soon transformed by collapse. The second main objection by \citet{COLLINS} was focused on the atmospheric dynamics, claiming that no metre-sized stony asteroid could survive down to the ground. Although \citet{GASP2} already shortly replied on this point, we would like to study better and to settle this latter issue. Aim of the present work is to evaluate, by means of order-of-magnitude calculations, if the hypothesis of Lake Cheko as an impact crater is consistent with the atmospheric dynamics and fragmentation of a cosmic body. 

\section{Uncertainties and unknowns in the input data}
Before studying the atmospheric fragmentation, we would like to underline some caveats in the input data. It is well known that by using a proper number of arbitrary parameters and by operating a fine tuning, every model could fit any data set. Therefore, the reliability of correctly designed models mostly depends on the quality of available input data\footnote{It is the well-known GIGO law: garbage in, garbage out.}. In the case of Tunguska, it is necessary to take into account many caveats. The impact occurred in an almost deserted region and eyewitness reports were from villages hundreds of kilometres away. Scientific instruments records were from observatories even farther: the closest observatory was at Irkutsk ($\sim 973$~km), while many barographs were placed in other European countries. Given such large distances, even a small percent error could have a significant impact on the parameters to be used in the models. For example, the epicentre calculated by \citep{BENMENAHEM} on the basis of seismograms was $\sim 5$~km distant from the more reliable epicentre calculated by Fast on the basis of fallen trees and confirmed by \citep{LONGO2}. This discrepancy is just $\sim 0.5$\% of the distance of calculated by using the data of the Irkutsk seismometer, and it is well explained by considering the differences in the terrain, which in turn could affect the transmission of seismic waves. However, this small discrepancy implies a difference in the direction to the Lake Cheko of $\sim 32^{\circ}$ with respect to the Fast's epicentre (Cheko-Ben Menahem's epicentre: azimuth $\sim 136^{\circ}$, distance $\sim 7$~km; Cheko-Fast's epicentre: azimuth $\sim 168^{\circ}$, distance $\sim 9$~km). Obviously, we considered the Fast's epicentre as the most reliable, but this example shows how small measurement errors could affect models input data. 

The environment of the region is the typical swampy Siberian taig\'a, with permafrost extending for metres below the ground, and seasonal floods generated by the ice thawing. The yearly floods imply significant changes in the environment. It is somehow surprising that the first scientific expedition led by Kulik in 1927, nineteen years after the TE, still found a relatively well preserved scenario of widespread devastation (fallen and burned trees). Perhaps, this triggered the thought that there still was the possibility to find evident craters and meteorites. Indeed, small holes were found, but the search for meteorites gave no result; later, it was suggested that these holes were due to the yearly thawing (see discussion in \citealt{GASP1}). It is reasonable to think that if there were relatively small craters or handful-sized stony or carbonaceous meteorites, they could have been covered and hidden by vegetation and mud of the yearly floods during the previous nineteen years (not to speak of today, after $111$~years!).

The impact that possibly generated the Lake Cheko occurred on a highly porous and soft terrain covered by trees and vegetation, crossed by a meander of the river Kimchu. As already suggested by \citep{GASP2}, an impact on such target should result in a transient crater, which in turn quickly collapse. The highly-porous terrain (cores revealed an average porosity of $75$\%, \citealt{GASP1,GASP3}) strongly absorb the impact energy (thus explaining the small depth-to-diameter ratio), hampering the formation of a rim and reducing the devastation of the nearby zone (cf \citealt{HOUSEN18,OKAMOTO} for impacts on porous materials). However, it is quite difficult to estimate the original crater shape, as the alluvial plain hosting Lake Cheko is a continuously changing environment, because of the yearly thawing and freezing, and variable inflow and outflow of the river Kimchu. Particularly, the yearly thawing results in the production of small circular holes on the terrain, with local small leakage of gas. Such holes were also visible at the lake bottom during many expeditions, as well as the scar of a recent landslide \citep{GASP4,GASP6}. This dynamic environment could justify local changes in the shape of the lake and its discrepancies with typical impact crater on solid dry ground. 

All these considerations stress the difficulty in finding reliable input parameters to reconstruct the TE. This means that models cannot conclusively demonstrate one hypothesis or another, but rather could give useful hints to reject unphysical hypotheses.

\section{Atmospheric Fragmentation}
Many theories were published on the atmospheric fragmentation of small asteroids/comets, some with specific application to the TE (just to cite a few: \citealt{OPIK,BALDWIN,SEKANINA1,TSVETKOV,CHYBA,HILLSGODA,SVETSOV,CEPLECHA,GRIGORIAN,KOROB,SEKANINA2,FOSCHINI1,BRONSHTEN,ARTESHU3,FOSCHINI2,SHUVALOV1,REVELLE,BOSLOUGH,POPOVA11,SHUVALOV2,COLLINS3,REGISTER,WHEELER,TABETAH}). In all of these theories, the key question is when fragmentation starts. The breakup of a solid body is essentially a random process: there is no known way to infer exactly the fragmentation threshold, and one could only derive statistical properties. In the case of a cosmic body entering a planetary atmosphere, the starting point was to consider as the threshold for breakup when the dynamical pressure in front of the cosmic body, traveling with speed $V$, exceeds its mechanical strength $S$ (e.g. \citealt{OPIK}): 

\begin{equation}
S \leq \rho_{\rm air} V^2
\label{eq:frag1}
\end{equation}

where $\rho_{\rm air}$ is the undisturbed air density. However, already \cite{OPIK} remarked that the breakup starts when the dynamical pressure at the stagnation point was one-two orders of magnitude smaller than the mechanical strength (see also \citealt{CEPLECHA96}; Table~2 of \citealt{FOSCHINI3}; Table~3 of \citealt{POPOVA11}). \cite{OPIK} suggested that the shear stress could be the main responsible of this effect. In addition, he noted that the cosmic body could either have internal cavities filled with gas or water, or be a rubble pile cemented with ice, something like the Whipple's dirty snowball. All these conditions could explain the breakup at dynamical pressures smaller than the mechanical strength. \cite{BALDWIN} and \cite{POPOVA11} proposed that the cosmic body could be cracked and flawed, because of past collisions occurred in space. The latter also suggested that the porosity could affect the fragmentation process. Also \cite{TSVETKOV} supported the hypothesis of a non-homogeneous structure of the asteroid. However, the general theory of shock compression and heating indicates that these processes eliminate the internal cracks and porosity, thus making the cosmic body more compact as it moves forward in the atmosphere \citep{ZELDOVICH}. There could still be internal voids so large to survive shock compression and thus altering the hypersonic flow \citep{FOSCHINI4,FOSCHINI2,TABETAH}, but this should be limited to a some cases only: as it depends on the internal structure and past history of the cosmic body, it should not be taken as a general rule.

In general, it is necessary to take into account that the scale-dependence of the mechanical strength follows a power law (e.g. \citealt{TURCOTTE,TSVETKOV,COTTOF}), as a particular case of the Weibull distribution \citep{WEIBULL}: 

\begin{equation}
\frac{S_{\rm i}}{S_{\rm f}} = \left( \frac{m_{\rm f}}{m_{\rm i}} \right)^{\alpha}
\label{eq:scalinglaw}
\end{equation}

where $m_{\rm i,f}$ are the initial and final masses, respectively, $S_{\rm i,f}$ are the corresponding mechanical strengths, and $\alpha$ is the scale factor (see \citealt{TURCOTTE} for some reference values). The larger is the body, the smaller is the strength. Therefore, by using the same mechanical strength both for meteorites and the original asteroid is not correct. Starting from laboratory measurements of meteorites strength, it is possible to extrapolate the breakup strength of the original cosmic body (Fig.~3 of \citealt{COTTOF} was based on the \citealt{POPOVA11} sample). Interestingly, it seems that larger boulders of either LL-chondrites ($\alpha \sim 0.16$) or carbonaceous chondrites ($\alpha \sim 0.2$) are more robust than ordinary H-chondrites ($\alpha \sim 2$). \cite{COTTOF} findings are valid within $2\sigma$ error, but they noted that there could be problems when dealing about rubble piles, as they could be a ``fluid'' aggregate of boulders. Rubble piles could also be composed by very different materials, with different densities (e.g. Itokawa asteroid, \citealt{MIYAMOTO}, \citealt{LOWRY}; $2008$~TC$_{3}$ Almahata Sitta meteorites, \citealt{BISCHOFF}). In addition, this evaluation can be done when meteorites are available (as in the case of Chelyabinsk, see the next Section); in the case of Tunguska, where no handful fragments were found, it is not possible to have a starting point to extrapolate the TCB mechanical strength.

\begin{table*}[t]
\caption{Comparison of fragmentation heights obtained in the present work with \cite{BOROVICKA} and \cite{POPOVA13}.}
\begin{center}
\begin{tabular}{lccc}
\hline
Quantity & \cite{BOROVICKA} & \cite{POPOVA13} & Present Work\\
\hline
Energy [kton] & $500$ &  $570-590$ & $561$\\
Velocity [km~s$^{-1}$] & $19.03$ & $19.16$ & 19.0\\ 
Mass [$10^7$~kg] & $1.2$ & $1.3$ & 1.3\\
Density [kg~m$^{-3}$] & 3300 & 3300 & 3300\\
Radius [m] & $9.5$ & $9.9$ & 9.8\\ 
Mechanical Strength [MPa] & 1 & 0.2 & 4\\
Fragmentation Height [km] & $45-30$ & $44-30$ & $40-32$\\
Fragmentation Height [km] & $26-24$ & $24$ & $33-26$\\
Fragmentation Height [km] & $22-20$ & $19$ & $28-21$\\
Major breakup [km] & $32-30$ & $30$ & $33-26$\\
\hline
\end{tabular}
\end{center}
\label{tab:chelyabinsk}
\end{table*}

Many other factors affect the atmospheric fragmentation of a cosmic body. In the following, we adopted the classical single-body theory, but with the fragmentation conditions by \cite{FOSCHINI1,FOSCHINI3}, which focuses on the hypersonic air flow around the cosmic body, and how it changes depending on the atmospheric conditions. The structure and properties of the Earth atmosphere (layers and pauses) depend on both the latitude and the season. Pauses are characterised by constant temperature as a function of the height, which implies a constant speed of sound ($V_{\rm sound}=\sqrt{\gamma R_{\rm air}T}$, with $R_{\rm air}$ specific gas constant for the air) and Mach number ($M=V/V_{\rm sound}$). Layers have changing temperature -- hence the speed of sound, and the Mach number -- as a function of the height. In the former case, the flow is steady, while it is unsteady in the latter. Therefore, the same type of cosmic body with a certain mass, strength, entry speed, and inclination angle could breakup at different height depending on the season and the latitude, as it encounters different pauses and layers, with steady and unsteady flows, respectively. There is not only one single fragmentation condition, but two. The first one, in the atmospheric pauses (steady flow), is when \citep{FOSCHINI1,FOSCHINI3}:

\begin{equation}
S \leq \frac{(1+ \iota)(\gamma - 1)}{2\gamma} \rho_{\rm air}V^2
\label{eq:vmax1}
\end{equation}

where $\gamma$ is the specific heat ratio and $\iota$ is the degree of ionisation of the flow. The ablation changes the properties of the flow at the stagnation point via the value of $\gamma$, which could range from $1.15$ to $3$ (see the discussion in \citealt{BRONSHTEN2,FOSCHINI1,FOSCHINI2}).

The second condition occurs in the atmospheric layers (unsteady flow): in this case, the interaction between shock waves and turbulence results in a sudden increase of the dynamical pressure up to more than one order of magnitude, depending also on the ionisation degree. Thus, the cosmic body begins to breakup when \citep{FOSCHINI3}:

\begin{equation}
S \leq \kappa (1+\iota)\rho_{\rm air}V^2
\label{eq:vmax2}
\end{equation}

where $2\lesssim \kappa \lesssim 6$ is the turbulence amplification factor. Fragmentation begins as one of the two above equations is satisfied first.  

The air density can be replaced by the equation linking the density at the sea level $\rho_0\sim 1.29$~kg~m$^{-3}$, the airburst height $h$ and the atmospheric scale height $H$:

\begin{equation}
\rho_{\rm air} = \rho_0 e^{-\frac{h+H}{H}}
\end{equation}

so to have a direct link between the height and the beginning of breakup. In the above equation, the fragmentation is considered to start at the height $(h+H)$ and to end after one scale height ($h$ is then the height when the fragmentation stops). It is worth reminding that also the atmospheric scale height depends on the temperature, and, in turn, on the season:

\begin{equation}
H=\frac{R_{\rm air}T}{g\cdot mol}
\end{equation}

where $g=9.81$~m~s$^{-2}$ is the Earth gravity acceleration and $mol\sim 0.029$~kg~mol$^{-1}$ is the mean molecular weight of the air.

As the cosmic body goes down in the Earth atmosphere (from $100$~km to the ground, $1$~km step size), the model first checks the steadiness/unsteadiness of the flow, and then applies the corresponding equation to determine the beginning of fragmentation. If there is breakup, the cosmic body is divided into boulders according to the selected distribution. The cycle restarts for each piece at the last calculated height and ends when the residual energy after the fragmentation is smaller than a certain value. If the program stops at height smaller or equal to the ground level, then there is an impact; otherwise, there is an airburst with complete vaporisation. 

\section{Model calibration with Chelyabinsk Event of 2013 February 15}
The Chelyabinsk Event (CE) of 2013 February 15 was the event with the greatest number of observations with different instruments and witnesses \citep{AVRAMENKO,BOROVICKA,BROWN,POPOVA13}. It happened at about $03^{\rm h}$:$20^{\rm m}$~UTC, corresponding at $09^{\rm h}$:$20^{\rm m}$ local time, over Chelyabinsk (Russia), a large city with about 1.2 millions inhabitants. Given the time and place, there were many eyewitnesses reports. Moreover, the wide diffusion of dash- and web-cams made it possible to have a large number of video registrations. Together with infrasonic, seismic, and satellite data, all these information made the CE the best asteroid airburst ever observed. We used this event to calibrate the general validity of our fragmentation model. For the atmospheric data, we downloaded the NRLMSISE-00 model\footnote{Available at \texttt{https://ccmc.gsfc.nasa.gov/}.} \citep{PICONE} for the coordinates and time of the event, starting from $100$~km height to the ground, with $1$~km step.

One key point is to assess the scaling laws for the cosmic body, that is the index $\alpha$ of Eq.~(\ref{eq:scalinglaw}). \cite{POPOVA13} reported about the measurements of mechanical strength of some gram-sized meteorites: C3, $4.46$~g, $330$~MPa; C4, $4.84$~g, $327$~MPa; C5, $1.58$~g, $408$~MPa. From these numbers is possible to estimate $\alpha\sim 0.2$. By assuming an initial mass of about $1.3\times 10^7$~kg, the extrapolated mechanical strength of the Chelyabinsk cosmic body results $\sim 4$~MPa, which is consistent with the estimates by \cite{BOROVICKA} and \cite{COTTOF}, while the $0.2$~MPa suggested by \cite{POPOVA13} seems to be too low. 

By combining the analyses of video records and sonic booms, \cite{BOROVICKA} estimated a series of major fragmentations between $40$ and $30$~km height, perhaps with mass loss already at $\sim 43$~km, followed by two more fragmentations at $24-26$~km and $20-22$~km (see Extended Data Figure 2 of \citealt{BOROVICKA}). The fragmentation heights are more or less similar to those estimated by \cite{POPOVA13}: main fragmentation between $44$ and $30$~km, and two more breakups at $\sim 24$ and $\sim 19$~km. The largest recovered boulder had a mass of $\sim 600$~kg, and its trajectory was deviated by $\sim 1.3^\circ$ (dispersion velocity $\sim 0.4$~km/s) falling in the Lake Chebarkul, at about $\sim 70$~km from the epicentre \citep{POPOVA13}.

In addition to atmospheric temperature and density as a function of height, the input parameters for our model refer to:

\begin{itemize}
\item cosmic body (energy released, geocentric velocity, inclination over the horizon, mechanical strength);
\item flow (degree of ionisation, specific heat ratio, turbulence amplification factor);
\item fragmentation (mass distribution of fragments, scaling factor for the mechanical strength).
\end{itemize}

In this case, the output values are the breakup heights to be compared with the observations. The results are displayed in Table~\ref{tab:chelyabinsk}. In our simulations, the final boulders are of the order of $200$~kg, a little underestimated, with a dispersion velocity of $\sim 0.3$~km/s estimated by using Eq.~(24) of \cite{PASSEY}. The latter equation is applied only when a fragment is smaller than $\sim 16$\% of the cosmic body size before the breakup, as  greater fragments could not exit from the hypersonic flow \citep{LAURENCE}.

\begin{figure}[t]
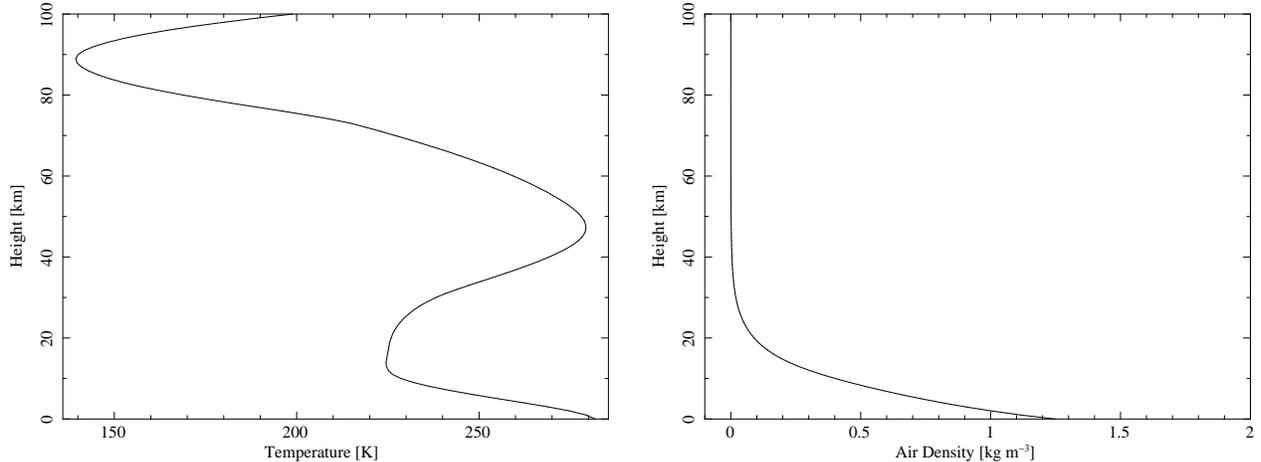

\begin{center}
\includegraphics[angle=270,scale=0.34]{NRL_temp}
\includegraphics[angle=270,scale=0.34]{NRL_dens}
\caption{Temperature (\emph{left panel}) and air density (\emph{right panel}) of the NRLMSISE-00 model for the coordinates of the TE \citep{PICONE}. }
\label{fig:atmodel}
\end{center}
\end{figure}

\section{Fragmentation of the Tunguska Cosmic Body}
After having validated our model, it is time to return back to the TE, trying to import some information gained with the CE simulations and comparison. We noted that the latitudes of the two events do not differ significantly: $\sim 61^\circ$~N for the TE, $\sim 55^\circ$~N for CE, but there is already an issue as TE was in summer, while CE in winter, which implies differences in the atmospheric temperature profiles. 
NRLMSISE-00 \citep{PICONE} model data starts from 1960, so we averaged the available values on June~30, $00^{\rm h}14'$~UTC, of each year, at the coordinates of the Tunguska Fast's epicentre (Latitude $60^{\circ}53'09''$~N, Longitude $101^{\circ}53'40''$~E, confirmed by recent reanalysis of \citealt{LONGO3}), starting from $100$~km height to the ground, with $1$~km step (Fig.~\ref{fig:atmodel}).

The lack of gram-size meteorites makes it impossible to estimate the mechanical strength of the TCB. Little is known about its composition: \cite{LONGO94}, by analysing microremnants found in the trees' resin, indicated the presence of Fe, Ca, Al, Si, Au, Cu, S, Zn, Cr, Ba, Ti, Ni, C, and O. \cite{ANFINOGENOV} suggested that the so-called ``John's Stone'', a large quarzitic rock found close to the epicentre of the TE, could be a piece of the TCB, but subsequent analyses of the oxygen isotopes made this hypothesis very unlikely \citep{BONATTI,HAACK}. Studies on the interplanetary dynamics of all possible orbits suggested an asteroidal origin \citep{SEKANINA1,SEKANINA2,ANDREEV,BRONSHTEN3,FARINELLA,JOPEK}, while models of atmospheric fragmentation exclude both the ice and the iron body (e.g. \citealt{CHYBA,HILLSGODA,FOSCHINI1}). To estimate the mechanical strength of the TCB, we can take into account these studies and guess some initial values around a few tens of megapascal. 

The resulting solutions are not consistent with the Lake Cheko scenario. To reach that region from the TE epicentre, a minimum deviation of $\sim 33^{\circ}$ with respect to TCB azimuth of $\sim 135^{\circ}$ \citep{LONGO2} is necessary. On one side, by a proper fine tuning of input parameters, metre-sized boulders could remains, but then they follow more or less the original trajectory with minimal deviations (just a few degrees). On the other side, if the final fragments have the necessary deviation, then their size is too small. We recall that, at Chelyabinsk, the largest boulder (about half ton) reached the Lake Chebarkul, about $70$~km West of the epicentre, but its deviation from the original azimuth was just $\sim 1.3^{\circ}$ \citep{BOROVICKA} or consistent with it \citep{POPOVA11}. 

The reason for the latter solution is due to the requirement to exit from the hypersonic flow: according to \cite{LAURENCE}, the fragment must be less than $0.16$ times the initial size of the cosmic body, otherwise it will remain trapped within the hypersonic flow. A TCB with radius of $\sim 30$~m could eject fragments smaller than $\sim 5$~m: this size is consistent with the Cheko crater, but -- as shown above -- the maximum attainable deviation from the original TCB azimuth is not enough. To set some order of magnitude, we recall the equation of the dispersion velocity of a fragment according to \cite{PASSEY}:

\begin{equation}
V_{\rm disp} = V\sqrt{\frac{3}{2}C\frac{r}{r_{\rm fr}}\frac{\rho_{\rm air}}{\delta}}
\label{eq:disp}
\end{equation}

where $V$ is the velocity of the cosmic body, $C\sim 0.02-1.52$ is an empirical factor based on the dispersion ellipses of meteorites, $r$ and $r_{\rm fr}$ are the radii of the initial cosmic body and of the fragment, respectively, $\rho_{\rm air}$ is the undisturbed air density, and $\delta$ is the cosmic body bulk density. In the case of TE, $\rho_{\rm air}\sim 0.521$~kg~m$^{-3}$ is the atmospheric density at $\sim 8$~km, and $r\sim 30$~m (by assuming a mass of $\sim 4\times 10^{8}$~kg and a density $\delta = 3300$~kg~m$^{-3}$). Then, by assuming the maximum value for $C$, it results from Eq.~(\ref{eq:disp}) that $V_{\rm disp}/V\sim 0.047$ for a $5$~m radius boulder. As the distance of the Lake Cheko from the Fast epicentre is $9$~km and $\sim 33^{\circ}$ with respect to the TCB azimuth of $\sim 135^{\circ}$ \citep{LONGO2}, the deviation from the TCB trajectory will be only $\sim 4^{\circ}$, not enough to reach the Lake Cheko. 

We can also perform the inverse calculation, to estimate the maximum fragment size that could have a dispersion velocity sufficient to deviate by $\sim 33^{\circ}$ from the TCB trajectory: it results that $r_{\rm fr}\lesssim 0.0023R$, which implies a fragment radius of $\sim 7$~cm ($\sim 5$~kg, by assuming $\delta=3300$~kg~m$^{-3}$). This is also in agreement with the numerical simulations by \cite{ARTESHU2}.

We would like to address also the possibility that TE was a less energetic airburst, as suggested by \cite{BOSLOUGH} on the basis of a hydrocode numerical simulation and by comparing with the forest distruction. According to their work, the energy released by the TE could have been smaller by a factor $2-3$ (i.e. $5$~Mton). This does not affect significantly the results of the present work: a smaller energy implies a smaller mass for the same entry velocity, which in turn means a slightly higher initial mechanical strength. The smaller mass results also in even smaller fragments to escape from the hypersonic flow: by assuming an entry speed of $\sim 16$~km/s, the estimated mass is $\sim 1.6\times 10^8$~kg, which in turn corresponds to a radius $\sim 22$~m ($\delta=3300$~kg~m$^{-3}$) and a maximum fragment size of $\sim 3.6$~m. Still enough to generate the Cheko crater, but the issue of the angle drift remains open, as Eq.~(\ref{eq:disp}) depends on the ratio between sizes.

\section{Binary Asteroid}
As the fallen trees pattern is consistent with the interaction of shocks from one couple of cosmic bodies, \cite{LONGO2,LONGO3} proposed that TCB could have been a binary asteroid: the main body vaporised at $\sim 8$~km from the ground, while its satellite continued and impacted one meander of the Kimchu river, excavating the Lake Cheko. 

Tidal disruptions (e.g. \citealt{WALSH}) or rotational fission (e.g. \citealt{PRAVEC}) could be two ways to generate a binary asteroid. A particular case of the tidal disruption is due to close encounters with planets. Given the relatively small size of the TCB and the need of the two bodies to remain close each other short before the impact, the latter seems to be the most viable option. Starting from the existence of a non-negligible number of double craters, \cite{FARINELLA92} proposed that close encounters of binary asteroids with the Earth could change their evolution, either by increasing or by decreasing their separation, and eventually leading to the impact with the planet. Later, \cite{BOTTKE2,BOTTKE1} applied the theory to rubble piles and found that Earth tidal forces could easily strip some fragment at each close encounter. The 1913 Chant meteor procession could be one of these case \citep{BEECH}. 

According to \cite{BOTTKE2,BOTTKE1}, the necessary characteristics of the rubble pile are: high prograde spin, close to the breakup limit; an elongated shape; periapsis close to the Earth; low geocentric speed; low bulk density. The size of the removed fragments is not necessarily comparable to the original asteroid: $10$\% is the most common value. In addition, planetary tides tend to align the main body and its fragment, so that the impacts are each close to the other. 

\begin{figure}[t]
\begin{center}
\includegraphics[angle=270,scale=0.35]{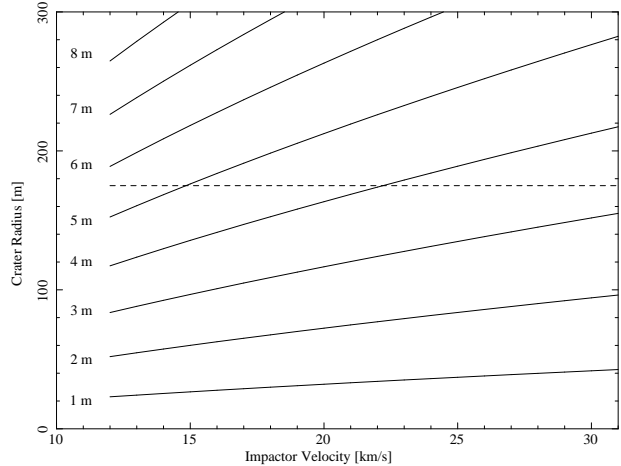}
\caption{Crater radius in wet soil as a function of the impactor radius and velocity ($\delta=3500$~kg~m$^{-3}$), according to values and formulae of Table~1 in \cite{HOLSAPPLE93}. The dashed line indicates the reference radius of $175$~m for the Lake Cheko.}
\label{fig:holsapple1993}
\end{center}
\end{figure}

\begin{figure}[ht]
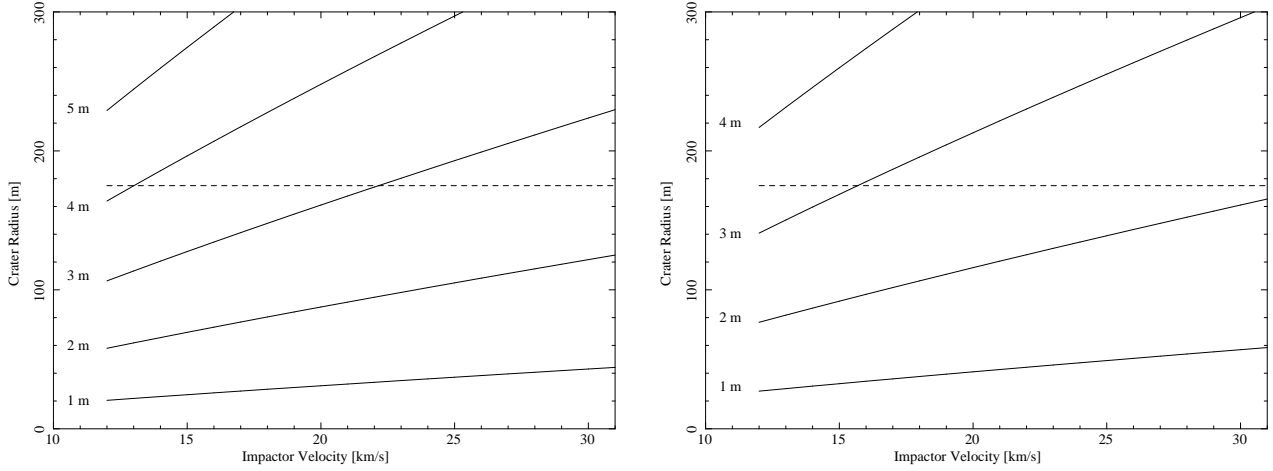

\begin{center}
\includegraphics[angle=270,scale=0.35]{Housen2018dens20}
\includegraphics[angle=270,scale=0.35]{Housen2018dens35}
\caption{Crater radius in porous soil ($\rho=1000$~kg~m$^{-3}$, $Y=1.0$~Mpa, porosity $75$\%) as a function of the impactor radius and velocity, according to \cite{HOUSEN18}. (\emph{left panel}) $\delta=2000$~kg~m$^{-3}$; (\emph{right panel}) $\delta=3500$~kg~m$^{-3}$. The dashed line indicates the reference radius of $175$~m for the Lake Cheko.}
\label{fig:housen2018dens2}
\end{center}
\end{figure}

According to \cite{GASP4}, there could be some metre-sized stony remnant (no indication of anomalous quantities of iron-nickel) on the bottom of the Lake Cheko, although it could simply be local porous terrain compressed by the impact (cf \citealt{HOUSEN18}). The terrain is highly porous ($75$\%, measured in the cores, \citealt{GASP1,GASP3}), with permafrost, and flooded because of thawing in June. To assess the impactor size, we calculated the crater size by using the equations by \cite{HOLSAPPLE93} for wet soil and by \cite{HOUSEN18} for porous terrain. The results are shown in Figs.~\ref{fig:holsapple1993} and \ref{fig:housen2018dens2}. By considering an impact velocity of about $15-20$~km/s, as for greater values the efficiency of tidal effects is smaller \citep{BOTTKE1}, the three analysed cases resulted in an impactor of about $3-5$~m radius to excavate a crater with $\sim 350$~m diameter and $\sim 50$~m depth.

One issue is to understand whether the secondary TCB experienced a fragmentation before impacting the ground. There is no indication of double airbursts in the seismograms and barograms, while the analysis of fallen trees is consistent with the existence of two or even more centres of explosions \citep{SERRA,LONGO2}. Recent simulations of ground effect of the Tunguska airburst suggest multiple airbursts, as, in the case of one single explosion, the North-Western part of the butterfly-shaped fallen tree region would have no damages \citep{NEMEC}. However, the main problem is the lack of any indication about this secondary airburst yield and height. In addition, the fragmentation strongly reduces the entry speed, which in turn implies the need of a larger body to excavate the same crater. For example, by considering the case in Fig.~\ref{fig:housen2018dens2}, right panel, but with an impact speed of $1$~km/s, a $\sim 26$~m body is required to excavate the Cheko crater. This means two unrealistic requirements: a significantly high mechanical strength of a fragment with such size ($\sim 10^8$~Pa), and a too large body size before the fragmentation.

Therefore, we focused on the possibility that the secondary TCB impacted the Kimchu region without fragmentation. The entry speed could remain high, thus implying a smaller, metre-sized body, for which an extreme compactness and high mechanical strength could be reasonable. According to \cite{COTTOF}, the greatest mechanical strength for a $3-5$~m radius cosmic body is of the order of $20$~MPa, in the case of a very low iron content chondrite (LL-type). However, the minimum mechanical strength required to avoid fragmentation is $\sim 300$~Mpa (by assuming the minimum turbulence amplification factor $\kappa=2$, and an extremely low ionisation degree $\iota=0.1$, and a cosmic velocity of $15$~km/s). To have such a mechanical strength in a metre-sized body, the scaling factor $\alpha$ should be equal to zero, which implies an extremely compact material, with no flaws, cracks or porosity. There was, at least, one interesting similar case happened in Carancas (Peru, September 15th, 2007): an ordinary H-chondrite with mass $\sim (7-12)\times 10^3$~kg (radius $\sim 0.8-1$~m), traveling at $12-17$~km/s, arrived without fragmentation to the ground and excavated an almost circular crater with $\sim 13$~m diameter on wet terrain \citep{TANCREDI}. According to \cite{COTTOF}, an H-chondrite of that size should have a strength of $\sim 0.1$~MPa. However, as the impact occurred at high-altitude ($\sim 3800$~m above the sea level), where the air density is $\rho_{\rm air}\sim 0.8$~kg~m$^{-3}$ (NRLMSISE00 model, \citealt{PICONE}), this implies that the cosmic body resisted to an uncorrected dynamic pressure $\rho_{\rm air}V^2\sim 115-231$~MPa. Therefore, Carancas is an example of how a high-strength high-compactness metre-sized stony objects could reach the ground.

Another option, which does not require a high-strength body, was suggested by \cite{GASP1} and \cite{LONGO3}: the impact could have caused thawing of the permafrost and releasing of water and trapped gas (such as methane), enlarging the original crater. Therefore, by taking into account these possible effects, the size of the impactor could be reduced. Obviously, a detailed analysis requires a hydrocode simulation, but once again we would like to write down some order-of-magnitude calculations to understand if the principle is valid.

The combustion of methane-air mixtures depends on a wide variety of pressure and temperature, with the best mixture having $\sim 10$\% in volume of methane \citep{KUNDU}. This is common in Siberia: just as a comparative example, the deep crater thawing-generated in Yamal (West Siberia) has a methane concentration in the nearby atmosphere of $\sim 10$\% \citep{LEIBMAN}. We consider a half-sphere of atmosphere surrounding the lake ($175$~m, volume $\sim 1.1\times 10^7$~m$^{3}$): the corresponding mass of methane at the local conditions of atmospheric pressure and temperature \citep{PICONE} is $\sim 7.5\times 10^{5}$~kg ($\rho_{\rm CH_{4}}=0.685$~kg~m$^{-3}$). The energy density released by methane combustion is $\sim 5\times 10^7$~J~kg$^{-1}$ \citep{WILLIAMS}, which implies that the combustion of the above cited methane mass releases an energy of $E_{\rm CH_{4}}\sim 3.7\times 10^{13}$~J. We compare with the minimum kinetic energy calculated above to excavate the lake Cheko: for example, we consider a stony cosmic body with $6$~m diameter ($\delta=3500$~kg~m$^{-3}$) and $15$~km~s$^{-1}$ entry velocity ($E_{\rm kin}\sim 4.5\times 10^{13}$~J). If we consider that the methane combustion is equivalent to add energy to the kinetic energy of a smaller body, then we remain with an energy of $E_{\rm kin}-E_{\rm CH_{4}}\sim 0.8\times 10^{13}$~J, which could be due to a cosmic body with $\sim 1.7$~m radius and the same entry velocity. Such a stony body has a much more common mechanical strength of the order of tens of megapascal \citep{COTTOF}, which allow it to hit the ground without fragmentation. 

\section{Final remarks}
We revisited the Tunguska event of June 30th, 1908, by taking into account the possibility that one metre-sized fragment impacted the ground forming the Lake Cheko. Our work favours the hypothesis that the TCB could have been a rubble-pile asteroid composed of very different materials with different strengths, density, and porosity. A close encounter with the Earth short before the impact with our planet could have divided into two pieces the TCB. The largest boulder ($\sim 60$~m diameter) produced the well-known airburst that devastated the Siberian taig\'a, while the smaller fragment ($6-10$~m diameter) continued without fragmentation toward the Kimchu river region and excavated the Lake Cheko. To reach the Kimchu river region without fragmentation, the secondary body was an extremely compact stone, with high mechanical strength ($\sim 300$~MPa). This is a high value, but not unrealistic, as shown by the 2007 Carancas meteorite case \citep{TANCREDI}. Another possibility to bypass the extreme compactness is to assume that the thawing of the permafrost released methane, which was ignited by the high temperatures developed in the impact, thus enlarging the original crater \citep{GASP1,LONGO3}. In this case, a $\sim 3.4-$m diameter boulder has an average mechanical strength and could reach the ground without fragmentation. We exclude the hypothesis of a single TCB, which ejected one metre-sized fragment during, or shortly before, the airburst: the lateral velocity for a metre-size boulder is not enough to deviate to reach the alleged impact site. Only centimetre-size fragment could have the required angular deviation. 

We would like to underline once again the speculative character of the present study, given the large uncertainties and unknowns in the input parameters for the atmospheric dynamics models. The present work is based on order-of-magnitude calculations, to understand if basic principles are well grounded, and, at least, we can conclude that some hypotheses are reliable. The burden of the proof remains in charge of those who will drill the Lake Cheko bottom, in search for some remnant of the TCB. This is the only way to write an end to this story.

\section{Acknowledgements}
LF acknowledges partial support by INAF Ricerca di Base 2017 and 2018. We thanks Enrico Bonatti and Giuseppe Longo for useful comments and suggestions.

\bibliography{tunguska}
\end{document}